\documentclass[11pt]{article}

\usepackage{amsmath}
\usepackage{graphicx}
\usepackage{amsfonts}
\usepackage{amssymb}
\usepackage{epsfig}
\usepackage{color}
\usepackage{psfrag}

\newcommand{\EQ}{\begin{equation}}
\newcommand{\EN}{\end{equation}}

\begin{document}
\setcounter{page}{0} \topmargin 0pt
\renewcommand{\thefootnote}{\arabic{footnote}}
\newpage
\setcounter{page}{0}

\begin{titlepage}

\vspace{0.5cm}
\begin{center}
{\Large {\bf Semiclassical Methods in 2D QFT: }}\\
{\Large {\bf  Spectra and Finite--Size Effects}}

\vspace{1cm}
{\large Valentina Riva} \\
\vspace{0.5cm} {\em School of Mathematics, Institute for Advanced Study}\\
{\em Einstein Drive, Princeton, NJ 08540, USA} \\
{\em riva@ias.edu}

\end{center}
\vspace{1cm}

\begin{abstract}
We review some recent results obtained in the analysis of
two--dimensional quantum field theories by means of semiclassical
techniques, which generalize methods introduced during the
Seventies by Dashen, Hasllacher and Neveu and by Goldstone and
Jackiw. The approach is best suited to deal with quantum field
theories characterized by a non--linear interaction potential with
different degenerate minima, that generates kink excitations of
large mass in the small coupling regime. Under these
circumstances, although the results obtained are based on a small
coupling assumption, they are nevertheless non--perturbative,
since the kink backgrounds around which the semiclassical
expansion is performed are non--perturbative too. We will discuss
the efficacy of the semiclassical method as a tool to control
analytically spectrum and finite--size effects in these theories.

\end{abstract}
\end{titlepage}

\section{Introduction}

Non--perturbative methods in quantum field theory (QFT) play a
central r\^ole in theoretical physics, with applications in many
areas, from string theory to condensed matter. During the last two
decades, considerable progress has been registered in the study of
two--dimensional systems, where exact results have been obtained
in the particular situations of conformally invariant or
integrable models (for detailed accounts of these achievements,
see for instance Refs.\cite{CFTbooks}, \cite{gmrep}).

A natural continuation of the above mentioned studies consists in
developing some techniques to control analytically
two--dimensional QFT which do not display conformal invariance or
integrability and therefore are presently analyzed by perturbative
or numerical methods only. This review summarizes the
contributions of the author in this respect, obtained in
collaboration with Giuseppe Mussardo and Galen Sotkov. Our main
tool was an appropriate generalization and extension of
semiclassical methods, which proved to be efficient in analysing
non-perturbative effects in QFT since their introduction in the
seminal works of Refs.\cite{DHN,GJ}. The semiclassical approach
does not require integrability, therefore it can be applied on a
large class of systems. At the same time, it permits to face
problems which are not fully understood even in the integrable
cases, such as the analytic study of QFT in finite volume. In
particular, it has led to new non--perturbative results on form
factors at a finite volume\cite{finvolff}, spectra of
non--integrable models\cite{dsgmrs} and energy levels of QFT on
finite geometries\cite{SGscaling,SGstrip,phi4}. Further
achievements in understanding advantages and drawbacks of the
semiclassical method have been presented by G. Mussardo in
Ref.\cite{giuseppe}.

The semiclassical method is best suited to deal with those quantum
field theories characterized by a non--linear interaction
potential with different degenerate minima. These systems display
kink excitations, associated to static classical backgrounds which
interpolate between neighbouring minima, which generally have a
large mass in the small coupling regime. Under these
circumstances, although the results obtained are based on a small
coupling assumption, they are nevertheless non-perturbative, since
the kink backgrounds around which the semiclassical expansion is
performed are non-perturbative too. The restriction on the variety
of examinable theories imposed by the above requirements is rather
mild, since non--linearity is the main feature of a wealth of
relevant physical problems.

The review is organized as follows. After recalling the basic
aspects of semiclassical quantization in
Section\,\ref{sectsemicl}, in Section\,\ref{sectnonint} we discuss
its application to the study of the spectrum in non--integrable
QFT. Section\,\ref{sectfinitesize} presents the analysis of
finite--size effects, and we conclude in
Section\,\ref{conclusions}.

\section{Semiclassical quantization}\label{sectsemicl}

In this Section we will describe the two main tools used in the
following to investigate non--integrable spectra and finite--size
effects. The first is represented by the semiclassical
quantization technique, introduced for relativistic field theories
in a series of papers by Dashen, Hasslacher and Neveu
(DHN)\cite{DHN} by using an appropriate generalization of the WKB
approximation in quantum mechanics. The second is a result due to
Goldstone and Jackiw\cite{GJ}, which relates the form factors of
the basic field between kink states to the Fourier transform of
the classical solution describing the kink. For a complete review
of these beautiful achievements, and complementary techniques
developed by other groups during the Seventies, see
Ref.\cite{raj}. Although semiclassical methods are naturally
formulated for QFT in any dimension $d+1$, here we will only
consider (1+1)--dimensional theories, in virtue of their
simplified kinematics that allows for powerful applications of the
semiclassical techniques.

\subsection{DHN method}\label{mainidea}

The semiclassical quantization of a field theory defined by a
Lagrangian
\begin{equation}\label{generallagr}
{\cal L}\,=\,\frac{1}{2}\left(\partial_\mu
\phi\right)\left(\partial^\mu \phi\right)-V(\phi)
\end{equation}
is based on the identification of a classical background
$\phi_{cl}(x,t)$ which satisfies the Euler--Lagrange equation of
motion
\begin{equation} \label{eom}
\partial_{\mu}\partial^{\mu}\phi_{cl}+V'(\phi_{cl})=0\;.
\end{equation}
The procedure is particularly simple and interesting if one
considers finite--energy static classical solutions $\phi_{cl}(x)$
in 1+1 dimensions, usually called \lq\lq kinks" or \lq\lq
solitons". They appear in field theories defined by a non--linear
interaction $V(\phi)$ displaying discrete degenerate minima
$\phi_{i}$, which are constant solutions of the equation of motion
and are called \lq\lq vacua". The (anti)kinks interpolate between
two next neighbouring minima of the potential, and they carry
topological charges $Q=\pm 1$.

Being static solutions of the equation of motion, i.e. time
independent in their rest frame, the kinks can be simply obtained
by integrating the first order differential equation related to
(\ref{eom})
\begin{equation}
\frac{1}{2}\left(\frac{\partial \phi_{cl}}{\partial x}\right)^{2}
\,=\, V(\phi_{cl}) + A \,\,\,, \label{firstorder}
\end{equation}
further imposing that $\phi_{cl}(x)$ reaches two different minima
of the potential at $x \rightarrow \pm \infty$. These boundary
conditions, which describe the infinite volume case, require the
vanishing of the integration constant $A$. As we will see in the
following, the kink solutions in a finite volume correspond
instead to a non--zero value of $A$, related to the size of the
system.

\begin{figure}[ht]
\begin{tabular}{p{5cm}p{5cm}p{5cm}}
\psfrag{phi}{$\phi$}\psfrag{V}{$V(\phi)$}\psfrag{b}{$\frac{m}{\sqrt{\lambda}}$}
\psfrag{a}{$\hspace{-0.3cm}-\frac{m}{\sqrt{\lambda}}$}
\psfig{figure=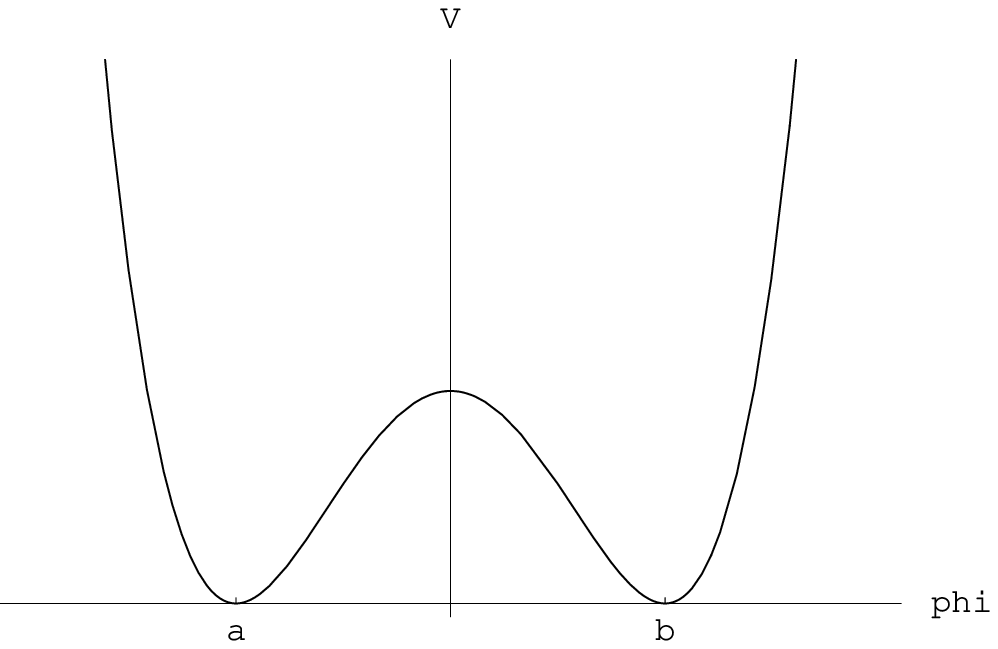,height=4cm,width=4.5cm} &
\psfrag{x}{$x$}\psfrag{phi}{$\phi_{cl}(x)$}\psfrag{c}{$\hspace{-0.4cm}\frac{m}{\sqrt{\lambda}}$}
\psfrag{d}{$\hspace{0.3cm}-\frac{m}{\sqrt{\lambda}}$}
\psfig{figure=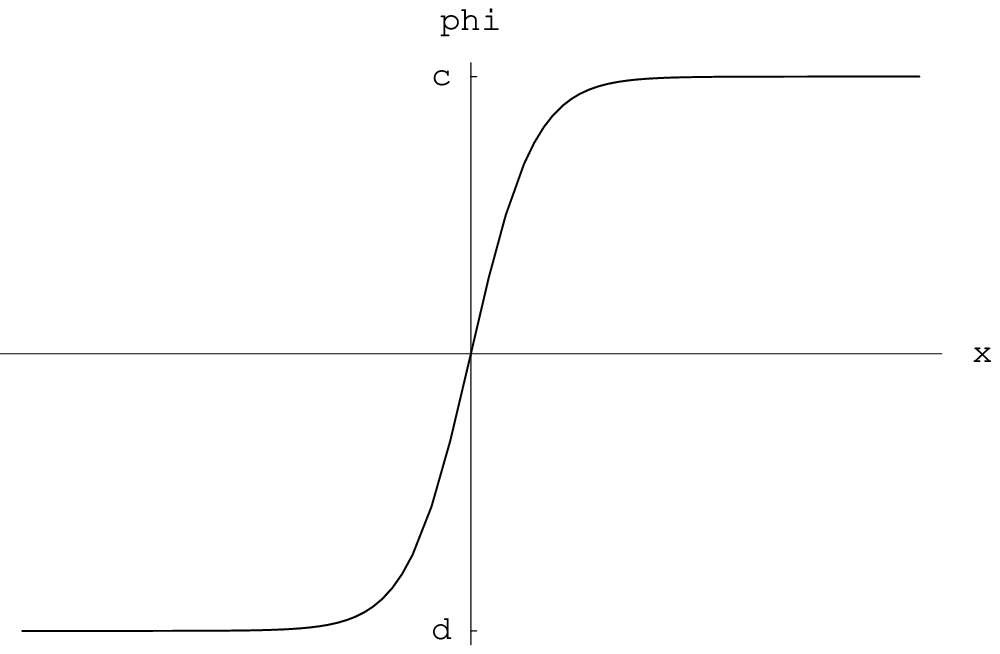,height=4cm,width=4.5cm} &
\psfrag{x}{$x$}\psfrag{ecl}{$\varepsilon_{cl}(x)$}
\psfig{figure=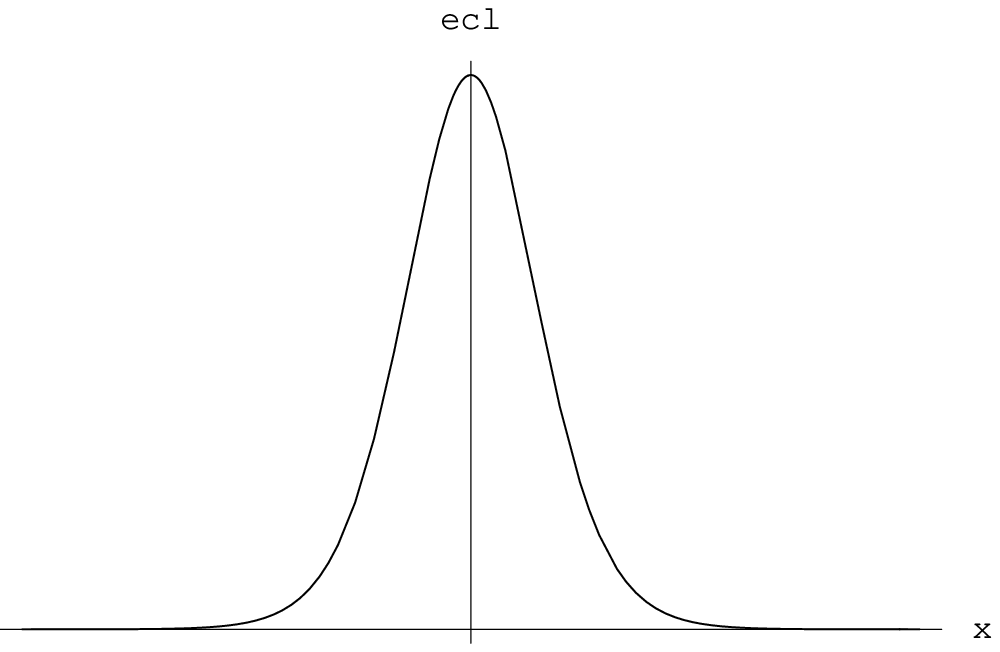,height=4cm,width=4.5cm}
\end{tabular}
\caption{Potential $V(\phi)$, kink $\phi_{cl}(x)$ and energy
density $\varepsilon_{cl}(x)$ in the broken $\phi^4$
theory.}\label{figphi4}
\end{figure}

For definiteness in the illustration of the method, we will focus
on the example of the $\phi^{4}$ theory in the broken
$\mathbb{Z}_2$ symmetry phase, defined by the potential
\begin{equation}\label{phi4pot}
V(\phi) = \frac{\lambda}{4}\,\phi^{4}-\frac{m^{2}}{2}\,\phi^{2} +
\frac{m^{4}}{4\lambda}\;.
\end{equation}
This theory displays two degenerate minima at
$\phi_{\pm}=\pm\frac{m}{\sqrt{\lambda}}$, and a static (anti)kink
interpolating between them
\begin{equation}\label{phi4kinkinf}
\phi_{cl}(x) \, = \,(\pm) \,\frac{m}{\sqrt{\lambda}}\;
\tanh\frac{m x}{\sqrt{2}}\;.
\end{equation}
The corresponding classical energy, obtained by integrating the
energy density $\varepsilon_{cl}(x)\equiv
\frac{1}{2}\left(\frac{d\phi_{cl}}{dx}\right)^2+V(\phi_{cl})\;$,
\begin{equation}\label{phi4kinkinfecl}
{\cal
E}_{cl}\equiv\int\limits_{-\infty}^{\infty}dx\;\varepsilon_{cl}(x)=\frac{2\sqrt{2}}{3}\frac{m^{3}}{\lambda}\;,
\end{equation}
diverges as the interaction coupling $\lambda\to 0$, indicating
that the solution is non--perturbative. Fig.\,\ref{figphi4} shows
the potential, the classical kink and its energy density.

At quantum level, the kinks are localized and topologically stable
excitations. An effective method for their semiclassical
quantization has been developed by Dashen, Hasslacher and Neveu
(DHN)\cite{DHN} by using an appropriate generalization of the WKB
approximation in quantum mechanics. The DHN method consists in
splitting the field $\phi(x,t)$ into the static classical solution
and its quantum fluctuations, i.e.
\begin{equation*}
\phi(x,t) \,=\,\phi_{cl}(x) + \eta(x,t) \,\,\, \,\,\, , \,\,\,
\,\,\, \eta(x,t) \,=\,\sum_{k} e^{i \omega_k t} \,\eta_{k}(x)
\,\,\,,
\end{equation*}
and in further expanding the action of the theory in powers of
$\eta$. This amount to an expansion in the interaction coupling
$\lambda$, as for instance in the example (\ref{phi4pot})
\begin{eqnarray}\nonumber
{\cal S}(\phi)&=&\int dx\,dt \,{\cal L}(\phi_{cl})+\int
dx\,dt\;\frac{1}{2}\,\eta(x,t)\left(\frac{d^{2}}{dt^{2}}-\frac{d^{2}}{dx^{2}}-m^{2}+3\lambda\phi_{cl}^{2}\right)\eta(x,t)
+\\ &&+\lambda\int
dx\,dt\,\left(\phi_{cl}\,\eta^{3}+\frac{1}{4}\eta^{4}\right)\;.\label{lagrexp}
\end{eqnarray}
The semiclassical approximation consists in keeping only the
quadratic terms in $\eta$. As a result of this procedure,
$\eta_{k}(x)$ satisfies the so called \lq\lq stability equation"
\begin{equation}\label{stability}
\left[-\frac{d^2}{d x^2} + V''(\phi_{cl}) \right] \, \eta_{k}(x)
\,=\, \omega_k^2 \,\eta_k(x) \,\,\,,
\end{equation}
together with certain boundary conditions. The semiclassical
energy levels in each sector are then built in terms of the energy
of the corresponding classical solution and the eigenvalues
$\omega_i$ of the Scr\"odinger--like equation (\ref{stability}),
i.e.
\begin{equation}
E_{\{n_i\}} \,=\,{\cal E}_{cl} + \hbar \,\sum_{k}\left(n_k +
\frac{1}{2}\right) \, \omega_k + O(\hbar^2) \,\,\,, \label{tower}
\end{equation}
where $n_k$ are non--negative integers. In particular the ground
state energy in each sector is obtained by choosing all $n_k = 0$
and it is therefore given by\footnote{From now on we will fix
$\hbar=1$, since the semiclassical expansion in $\hbar$ is
equivalent to the expansion in the interaction coupling
$\lambda$.}
\begin{equation}
E_{0} \,=\,{\cal E}_{cl} + \frac{\hbar}{2} \,\sum_{k} \omega_k +
O(\hbar^2) \,\,\,. \label{e0}
\end{equation}

This technique was applied in Ref.\cite{DHN} to the kink
background (\ref{phi4kinkinf}), in order to compute the first
quantum corrections to its mass, whose leading order term is the
classical energy. In this case, the stability equation
(\ref{stability}) can be cast in hypergeometric form and therefore
exactly solved. The semiclassical correction to the kink mass can
be computed as the difference between the ground state energy in
the kink sector and the one of the vacuum sector, plus a mass
counterterm due to normal ordering (see Ref.\cite{raj} for
details). The final result is
\begin{equation}\label{kinkmassoneloop}
M\,=\,\frac{2\sqrt{2}}{3}\,\frac{m^{3}}{\lambda}+m\left(\frac{1}{6}\,\sqrt{\frac{3}{2}}-\frac{3}{\pi\sqrt{2}}\right)\;.
\end{equation}

\subsection{Classical solutions and form factors}\label{secsemFF}

A direct relation between the kink states and the corresponding
classical solutions has been established by Goldstone and
Jackiw\cite{GJ}, who have shown that the matrix element of the
field $\phi$ between kink states is given, at leading order in the
semiclassical expansion, by the Fourier transform of the kink
background.

We will now derive this result in the example of the broken
$\phi^{4}$ theory (\ref{phi4pot}), illustrating the assumptions
behind it. Let us define the matrix element (also called form
factor) of the basic field $\phi(x,t)$ between two one--kink
states of momenta $p_{1}$ and $p_{2}$, as the Fourier transform of
a function $\hat{f}(a)$, to be determined:
\begin{equation}
\langle\,p_{2}\,|\,\phi(0)\,|\,p_{1}\,\rangle\;=\,\int
da\;e^{i(p_{1}-p_{2})a}\;\hat{f}(a)\;.
\end{equation}
Next consider the Heisenberg equation of motion for the quantum
field $\phi(x,t)$
\begin{equation}
\left(\partial_{t}^{2}-\partial_{x}^{2}\right)\phi(x,t)\,=\,m^{2}\,\phi(x,t)\,-\,\lambda\,\phi^{3}(x,t)\;,
\end{equation}
and take the matrix elements of both sides\footnote{Lorentz
invariance imposes the relation
$\langle\,p_{2}\,|\,\phi(x,t)\,|\,p_{1}\,\rangle\,=\,
e^{-i(p_{1}-p_{2})_{\mu}x^{\mu}}\langle\,p_{2}\,|\,\phi(0)\,|\,p_{1}\,\rangle$}
\begin{equation*}
\left[-(p_{1}-p_{2})_{\mu}(p_{1}-p_{2})^{\mu}\right]
e^{-i(p_{1}-p_{2})_{\mu}x^{\mu}}\langle\,p_{2}\,|\,\phi(0)\,|\,p_{1}\,\rangle
\,=\,
\end{equation*}
\begin{equation}\label{heis}
=\,e^{-i(p_{1}-p_{2})_{\mu}x^{\mu}}\,\left\{\,m^{2}\,\langle\,p_{2}\,|\,\phi(0)\,|\,p_{1}\,\rangle
\,-\,\lambda\,\langle\,p_{2}\,|\,\phi^{3}(0)\,|\,p_{1}\,\rangle\,\right\}\;.
\end{equation}
We will now equate the two members of this equation at leading
order in $\lambda$.

In the left hand side of (\ref{heis}), the energy difference
$$
\left(E_{1}-E_{2}\right)^{2}=\left(\frac{p_{1}^{2}-p_{2}^{2}}{2M}+...\right)^{2}=O(\lambda^{2})
$$
can be neglected, since the kink momentum is very small compared
to its mass, due to
(\ref{phi4kinkinfecl}),(\ref{kinkmassoneloop}). Hence the left
hand side gives, at leading order,
$$
\int
da\;e^{i(p_{1}-p_{2})a}\;\left(-\frac{d^{2}}{da^{2}}\,\hat{f}(a)\right)\;.
$$

In the right hand side of (\ref{heis}), the cubic power
$\langle\,p_{2}\,|\,\phi^{3}(0)\,|\,p_{1}\,\rangle$ can be
expanded over a complete set of states with the same topological
charge as the kink. These are given by one--kink states
$|\,p\,\rangle$ and by kink + neutral states
$|\,p,k_1,...,k_m\,\rangle$, where $k_i$ indicate the neutral
states' momenta (the neutral states, also called "mesons", are the
quantum excitations associated to constant classical backgrounds,
i.e. to vacua). Our assumption is that, in the weak coupling
limit, the corresponding matrix elements behave as
$\langle\,p'\,|\,\phi(0)\,|\,p\,\rangle=O(1/\sqrt{\lambda})$ and
$\langle\,p',k'_1,...,k'_l\,|\,\phi(0)\,|\,p,k_1,...,k_m\,\rangle=O(\lambda^{(l+m-1)/2})$.
This assumption, which will find confirmation \textit{a
posteriori}, relies on the fact that the kink classical background
is of order $1/\sqrt{\lambda}$ itself, and that the emission or
absorption of every meson carry a factor $\sqrt{\lambda}$
\footnote{This can be intuitively understood by noticing that, in
the expansion (\ref{lagrexp}) of the interaction $V(\phi)$, the
leading perturbative term is of order $\lambda\phi_{cl}$, i.e. of
order $\sqrt{\lambda}$.}. In virtue this assumption, the leading
term is obtained when the intermediate states are all one--kink
states:
\begin{equation*}\label{intermediate}
-\lambda\sum_{p,q}
\langle\,p_{2}\,|\,\phi(0)\,|\,p\,\rangle\langle\,p\,|\,\phi(0)\,|\,q\,\rangle\langle
\,q\,|\,\phi(0)\,|\,p_{1}\,\rangle\,=\,-\lambda \int
da\;e^{i(p_{1}-p_{2})a}\;\,[\,\hat{f}(a)\,]^{3}\;.
\end{equation*}
Hence, at leading order in $\lambda$, the function $\hat{f}(a)$
obeys the same differential equation satisfied by the kink
solution, i.e.
\begin{equation}
\frac{d^{2}}{da^{2}}\hat{f}(a)=\lambda[\,\hat{f}(a)\,]^{3}-m^{2}\hat{f}(a)\;.
\end{equation}
This means that we can take $\hat{f}(a)  = \phi_{cl}(a)$,
adjusting its boundary conditions by an appropriate choice for the
value of the constant $A$ in eq.\,(\ref{firstorder}).

Therefore, we finally obtain
\begin{equation}\label{finalGJ}
\langle\,p_{2}\,|\,\phi(0)\,|\,p_{1}\,\rangle\;=\,\int
da\;e^{i(p_{1}-p_{2})a}\;\phi_{cl}(a)\;+\;\text{higher order
terms}\;.
\end{equation}
Along the same lines, it is easy to prove that the form factor of
an operator expressible as a function of $\phi$ is given by the
Fourier transform of the same function of $\phi_{cl}$. For
instance, the form factor of the energy density operator
$\varepsilon$ can be computed performing the Fourier transform of
$\varepsilon_{cl}(x) =
\frac{1}{2}\left(\frac{d\phi_{cl}}{dx}\right)^{2} +
V\left(\phi_{cl}\right)$.

Similar arguments lead to semiclassical expression for the matrix
elements between excited states of the kink, or states containing
kink and mesons (see Ref.\cite{GJ},\cite{raj}).

\section{Non--integrable quantum field theories}\label{sectnonint}

As we mentioned in the Introduction, integrable QFT in $(1+1)$
dimensions admit a non--perturbative treatment which led to exact
results and relevant predictions in statistical mechanics and
condensed matter applications. Nonetheless, both theoretical
reasons and applications call for a deeper understanding of
non--integrable systems as well. In general, these are analyzed
through perturbative or numerical techniques only, and some of
their basic data, such as the mass spectrum, are often not easily
available. There are, however, two favorable situations when
analytical tools can be used to extract non--perturbative results.

The first case is that of non--integrable theories which can be
seen as small perturbations of integrable ones. An approach called
Form Factor Perturbation Theory (FFPT)\cite{DMS} has been
developed, which exploits the non--perturbative knowledge of the
integrable theory in order to get quantitative predictions on mass
spectrum, scattering amplitudes and other physical quantities in
these systems.

A complementary situation is represented by theories having kink
excitations of large mass in their semiclassical limit. In this
case, the semiclassical method introduced in
Section\,\ref{sectsemicl} is a natural candidate to obtain
analytic non--perturbative results. In this Section we focus on
this approach, and we apply it to the analysis of the spectrum in
some non--integrable theories. Our main tool will be a
generalization of the result by Goldstone and Jackiw discussed in
Sect.\,\ref{secsemFF}.

\subsection{Relativistic formulation of Goldstone and Jackiw's
result}\label{sectGJimprov}

In order to apply Goldstone and Jackiw's result (\ref{finalGJ}) to
the study of the spectrum in non--integrable QFT, we first need to
refine it in order to overcome its drawback of being expressed in
terms of the difference of space momenta of the two kinks. The
original formulation is not Lorentz covariant, and the
antisymmetry under the interchange of momenta makes problematic
any attempt to go in the crossed channel and obtain the matrix
element between the vacuum and a kink--antikink state.

\begin{figure}[h]
\psfrag{phi}{$\hspace{0.15cm}\phi$}\psfrag{K}{$K$}\psfrag{Kb}{$\bar{K}$}
\psfrag{b}{$b$}\psfrag{Gabc}{$\Gamma_{K\bar{K}}^{b}$}\psfrag{(a)}{$(a)$}\psfrag{(b)}{$(b)$}\psfrag{(c)}{$(c)$}
\hspace{1cm}\psfig{figure=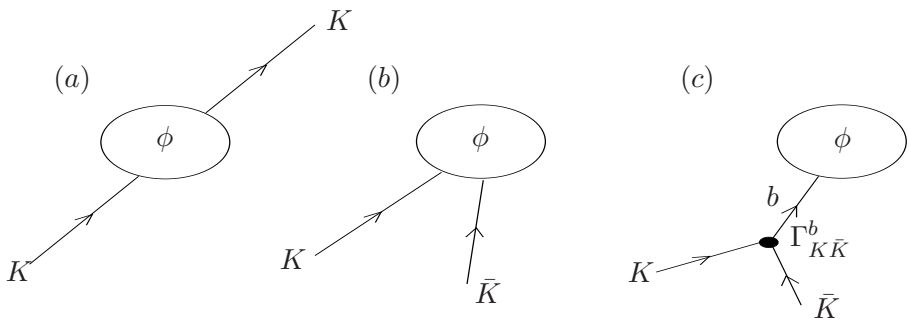,height=4cm,width=12cm}
\caption{Pictorial representation of $(a)$ the form factor
$f(\theta)$, $(b)$ the crossed channel form factor $F_2(\theta)$,
$(c)$ the form factor $F_2(\theta^*)$ at the dynamical pole
$\theta^{*}$.} \label{figGJref}
\end{figure}

In order to overcome these problems, in Ref.\cite{finvolff} we
have refined the result by using, instead of the space--momenta of
the kinks, their rapidity variable $\theta$, defined in terms of
energy and momentum as
\begin{equation}\label{rapiditydef}
E\equiv M \cosh\theta\;,\qquad p\equiv M \sinh\theta\;.
\end{equation}
This parameterization is particularly convenient, since the
rapidity difference is a Lorentz invariant of a two--particles
scattering process, as can be seen from its relation with the
Mandelstan variable $s$:
\begin{equation}\label{softheta}
s=(p_1+p_2)_{\mu}(p_1+p_2)^{\mu}=m_1^2+m_2^2+2m_1 m_2
\cosh(\theta_1-\theta_2)\;.
\end{equation}
The approximation of large kink mass used by Goldstone and Jackiw
can be realized considering the rapidity as very small. For
example, in the $\phi^{4}$ theory (\ref{phi4pot}), where the kink
mass $M$ is of order $1/\lambda$, we work under the hypothesis
that $\theta$ is of order $\lambda$. In this way we get $E\simeq
M\;$,$ \;p\simeq M\,\theta \ll M$. It is easy to see that the
proof of (\ref{finalGJ}) outlined in Sect.\,\ref{secsemFF} still
holds, if we define the form factor between kink states as the
Fourier transform with respect to the rapidity difference
$\theta=\theta_1-\theta_2$
\begin{equation}\label{ffinf}
\langle \,p_{1}|\,\phi(0)\,|p_{2}\,\rangle\equiv f(\theta)\equiv
M\int da\,e^{i\,M\theta a}\phi_{cl}(a)\;.
\end{equation}
The use of rapidity variable permits to analytically continue the
form factor (\ref{ffinf}) to the crossed channel, via the
transformation $\theta\rightarrow i\pi-\theta$, which is
equivalent to the transformation from the Mandelstam variable $s$
to $t$. We then have
\begin{equation}\label{f2}
F_{K\bar{K}}(\theta) \,\equiv\,
\langle\,0|\,\phi(0)|\,p_{1},\bar{p}_{2}\,\rangle =
f(i\pi-\theta)\;.
\end{equation}
A first use of the matrix elements (\ref{f2}) is to estimate the
leading behaviour in $\lambda$ of the spectral representation of
correlation functions in a regime of momenta dictated by the
assumption of small kink rapidity. Here we will discuss their
second main application, which permits to extract information
about the spectrum of the theory. The two--particle form factors
share the same $s$--channel dynamical poles of the scattering
matrix, which correspond to the creation of kink--antikink bound
states. Their behaviour in the vicinity of a singularity is
$$
F_{K\bar{K}}(\theta)\,\sim\,\frac{i\,\Gamma_{K\bar{K}}^{b}}{(\theta-\theta^{*})}\,F_{b}(\theta^*)\;,
$$
where $\Gamma_{K\bar{K}}^{b}$ is the on--shell three--particle
coupling constant between kink, antikink and the bound state $b$,
and the poles are located at $\theta^{*}=i(\pi-u)$, with
$0<u<\pi$. The process is pictorially represented in
Fig.\,\ref{figGJref}. Since the corresponding singularity in the
$s$--variable is of the form $(s-m^2_b)^{-1}$, it follows from
(\ref{softheta}) that the mass of the bound state can be expressed
as
\begin{equation}
m_b^2\,=\,m_K^2+m^2_{\bar{K}}+2m_Km_{\bar{K}}\cos u \,= \, \left(2
M \sin\frac{u}{2}\right)^2\;.
\end{equation}

It is worth noticing that this procedure for extracting the
semiclassical bound states masses is remarkably simpler than the
standard DHN method of quantizing the corresponding classical
backgrounds, because in general these solutions depend also on
time and have a much more complicated structure than the kink
ones. Moreover, in non--integrable theories these backgrounds
could even not exist as exact solutions of the field equations:
this happens for example in the $\phi^{4}$ theory, where the DHN
quantization has been performed on some approximate
backgrounds\cite{DHN}.

\subsection{Broken $\phi^4$ theory}

Let us now apply the semiclassical method to the analysis of the
spectrum in the $\phi^4$ field theory in the $\mathbb{Z}_2$ broken
symmetry phase. This non--integrable theory, defined by the
potential (\ref{phi4pot}), is invariably referred to as a paradigm
for a wealth of physical phenomena. In spite of this deep
interest, however, its non--perturbative features are still poorly
understood.

The main properties of the potential (\ref{phi4pot}) and its kink
background (\ref{phi4kinkinf}) have been already discussed in
Sect.\,\ref{mainidea}. The form factor (\ref{ffinf}) takes the
form
\begin{equation}
\langle\,p_{2}|\,\phi(0)|\,p_{1}\,\rangle \, =\, \frac{4}{3}i\pi
\left(\frac{m}{\sqrt{\lambda}}\right)^{3}
\frac{1}{\sinh\left(\frac{2}{3}\pi\frac{m^{2}}{\lambda}\,\theta\right)}\;,
\label{phi4ffinfvol}
\end{equation}
where the kink mass is approximated at leading order by the
classical energy $M=\frac{2\sqrt{2}}{3}\frac{m^{3}}{\lambda}$. The
dynamical poles of $F_{K\bar{K}}(\theta)$ are located at
\begin{equation}\label{noflambda}
\theta_{n}=i\pi\left[1-\frac{3}{2\pi}\,\frac{\lambda}{m^{2}}\,n\right]\,,
\quad\qquad 0<n<\frac{2\pi}{3}\frac{m^{2}}{\lambda} \,,
\end{equation}
and the corresponding bound states masses are given by
\begin{equation}\label{phi4boundst}
m_{b}^{(n)} = 2M\sin\left[\frac{3}{4}\,
\frac{\lambda}{m^{2}}\,n\right] =
n\,\sqrt{2}\,m\left[1-\frac{3}{32}\,
\frac{\lambda^{2}}{m^{4}}\,n^{2}+...\right]\,.
\end{equation}
Note that the leading term is consistently given by multiples of
$\sqrt{2}m$, which is the known mass of the elementary boson of
the theory\footnote{The elementary bosons represent the
excitations over the vacua, i.e. the constant backgrounds
$\phi_{\pm}=\pm\frac{m}{\sqrt{\lambda}}$, therefore their square
mass is given by $V''(\phi_{\pm})=2\,m^2$.}. This spectrum exactly
coincides with the one derived in Ref.\cite{DHN} by building
approximate time--dependent classical solutions to represent the
neutral excitations. Since $m_b^{(3)}>2m_b^{(1)}$, even when more
than two particles are allowed by the value of $\lambda$ in
(\ref{noflambda}), only the first two are stable, while the others
are resonances (see Ref.\cite{giuseppe} for further comments and
generalizations).

\subsection{Efficacy and limitations of the semiclassical method}

We have seen in the previous Section that the semiclassical method
is an efficient tool to get the spectrum of the $\phi^4$ theory.
The natural question to be addressed now is how reliable the
method is to study other types of potential.

\begin{figure}[ht]
\begin{tabular}{p{5cm}p{5cm}p{5cm}}
\psfrag{V}{$V$}\psfrag{phi}{$\phi$}\psfrag{a}{$$}\psfrag{b}{$0$}
\psfrag{c}{$$}\psfrag{d}{$\frac{2\pi}{\beta}$}\psfrag{e}{$$}\psfrag{f}{$\frac{4\pi}{\beta}$}
\psfig{figure=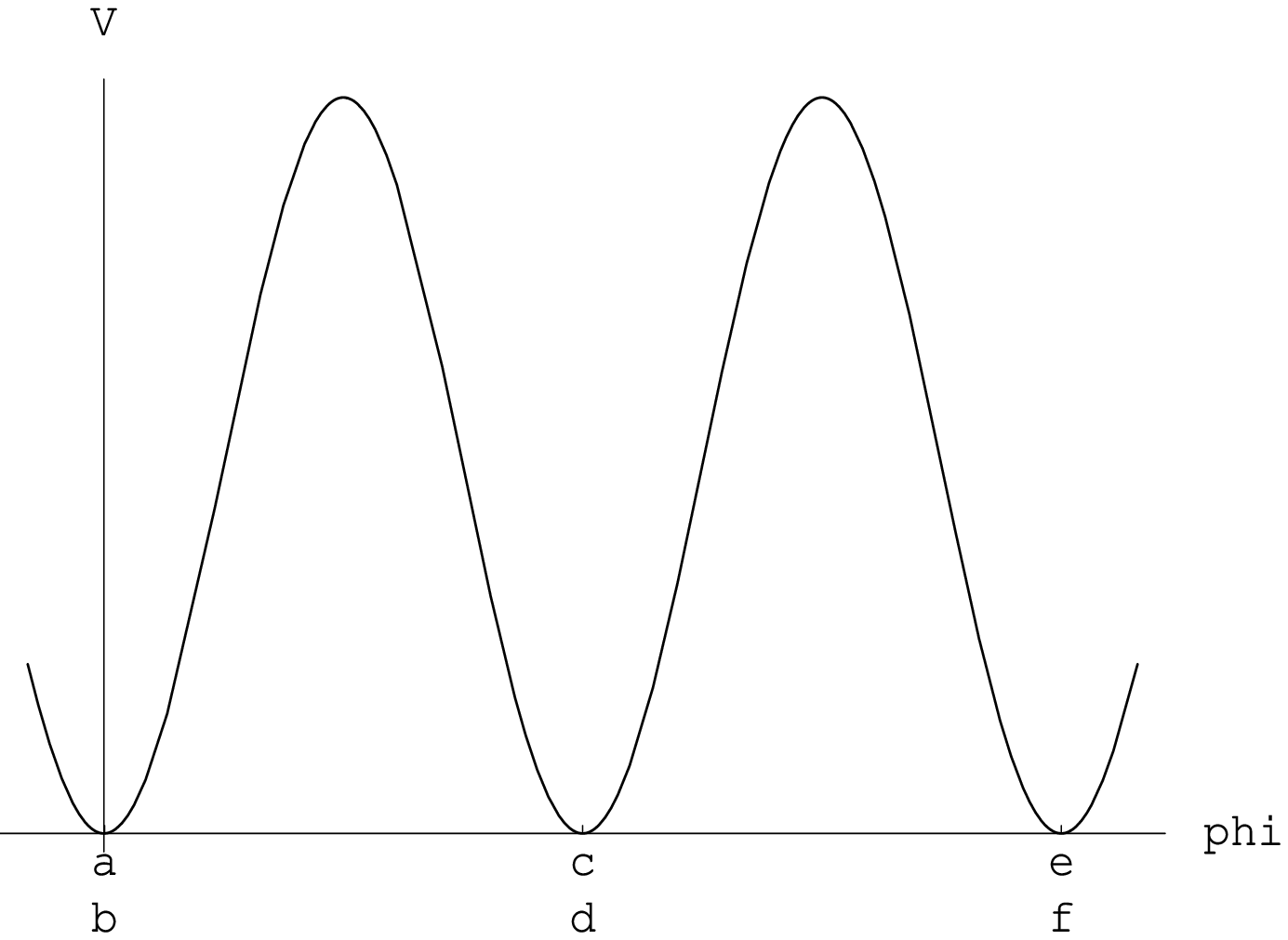,height=4cm,width=4.5cm} &
\psfrag{V}{$V$}\psfrag{phi}{$\phi$}\psfrag{a}{$$}\psfrag{b}{$0$}
\psfrag{c}{$$}\psfrag{d}{$\frac{2\pi}{\beta}$}\psfrag{e}{$$}\psfrag{f}{$\frac{4\pi}{\beta}$}
\psfig{figure=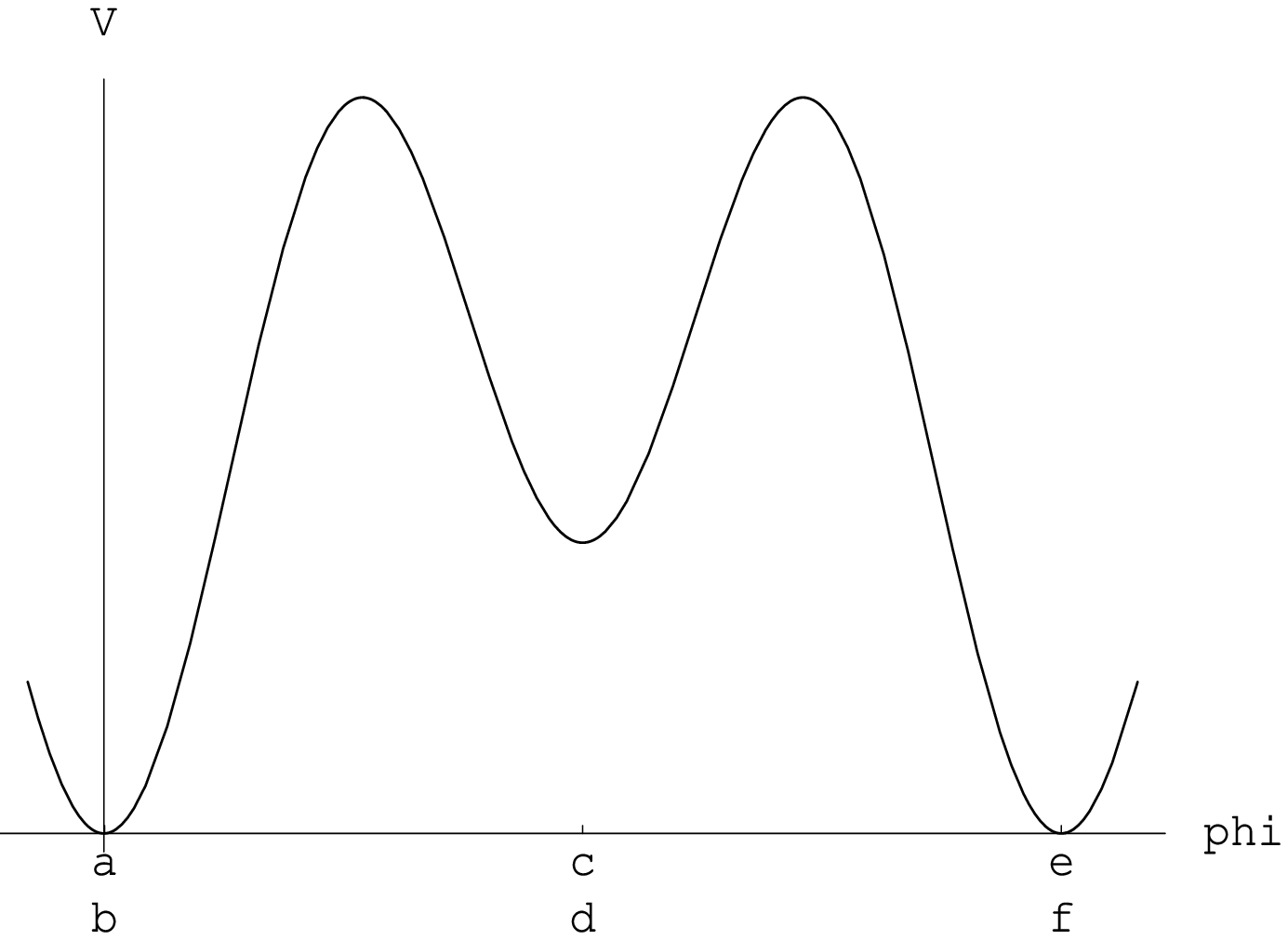,height=4cm,width=4.5cm} &
\psfrag{V}{$V$}\psfrag{phi}{$\phi$}\psfrag{a}{$$}\psfrag{b}{$0$}
\psfrag{c}{$$}\psfrag{d}{$\frac{2\pi}{\beta}$}\psfrag{e}{$$}\psfrag{f}{$\frac{4\pi}{\beta}$}
\psfig{figure=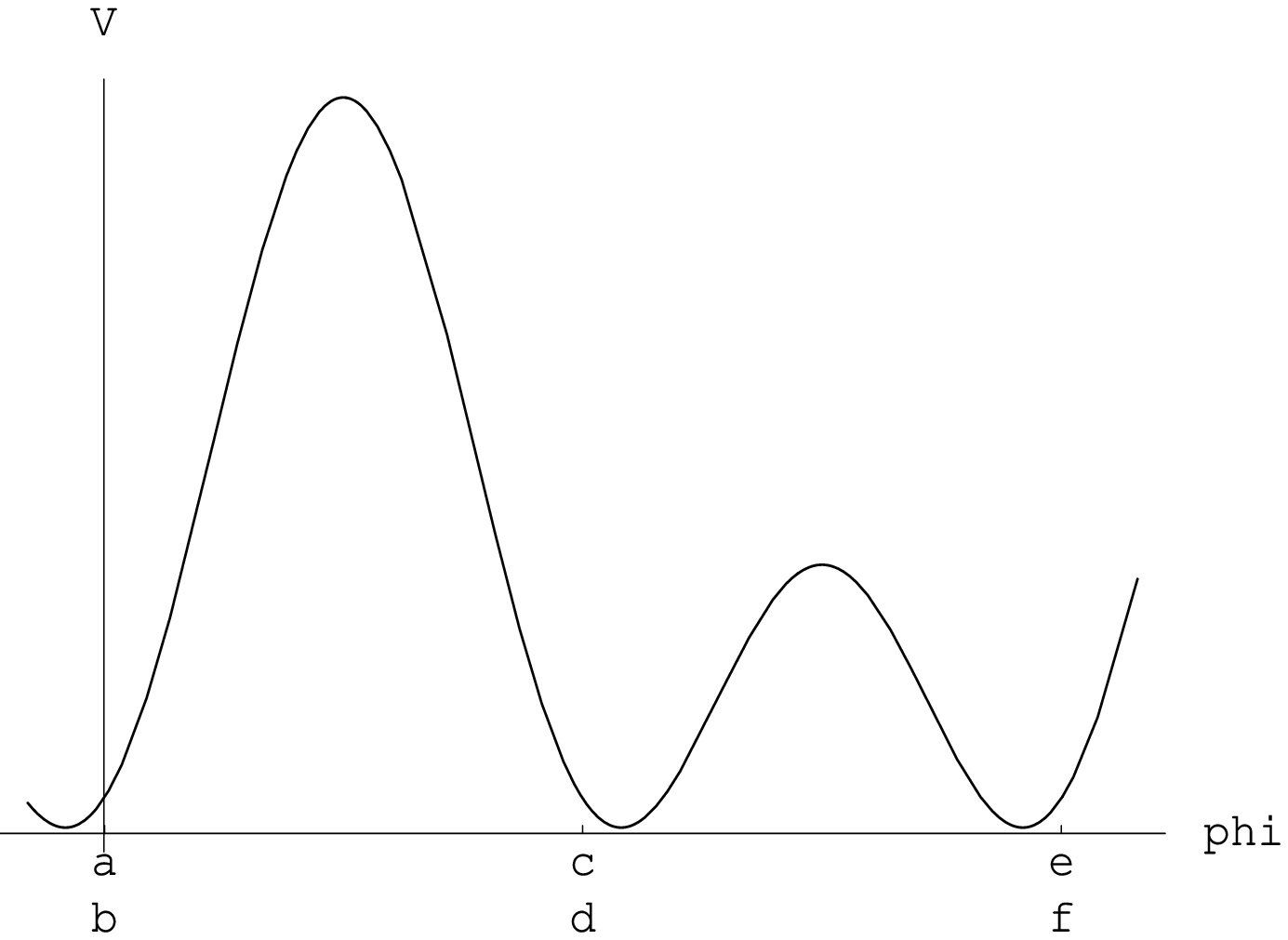,height=4cm,width=4.5cm}
\end{tabular}
\caption{Potential $V(\phi)$ defined in (\ref{VSG}), in
(\ref{VDSG}) with $\delta=0$ and in (\ref{VDSG}) with
$\delta=\frac{\pi}{2}$, respectively.}\label{figSG}
\end{figure}

First, let us mention an example where the semiclassical results
can be compared, in the appropriate regime of couplings, with
exact results obtained in virtue of
integrability\cite{SGintegrable}. This is the case of the
Sine--Gordon model, defined by the potential
\begin{equation}\label{VSG}
V(\phi)\,=\,\frac{m^2}{\beta^{2}}\,(1-\cos\beta\phi)
\end{equation}
(see the first picture in Fig.\,\ref{figSG}). For this model, the
semiclassical results are in very good agreement with exact ones
also for values of $\beta$ which extend beyond the semiclassical
limit (see Refs.\cite{SGscaling},\cite{giuseppe} for details). The
spectrum consists of soliton and antisoliton excitations, which
classically interpolate between two neighbouring minima of
(\ref{VSG}), and a tower of neutral states, called "breathers",
associated to every minima.

A very interesting non--integrable theory is defined by the Double
Sine--Gordon potential
\begin{equation}\label{VDSG}
V(\phi)\,=\,\frac{m^2}{\beta^{2}}\,(1-\cos\beta\phi)\,+\,\frac{\lambda}{\beta^2}\,\cos\left(\frac{\beta}{2}\phi+
\delta\right)\,+\,const\;,
\end{equation}
(see Fig.\,\ref{figSG}) which has several applications in
statistical mechanics and condensed matter physics.

From the theoretical point of view, this model is an ideal
laboratory where to compare results obtained by FFPT around the
integrable Sine--Gordon model\cite{dm,hungDSG} (for small
$\lambda$) and semiclassical methods\cite{dsgmrs} (for small
$\beta$). Depending on the value of the phase $\delta$, the
potential (\ref{VDSG}) displays different qualitative features,
which can be grouped in two classes: $\delta\neq\frac{\pi}{2}$ and
$\delta=\frac{\pi}{2}$. We will now briefly summarize the results
obtained in Ref.\cite{dsgmrs}, to which we refer the reader for
details.

Let us first describe the  $\delta\neq\frac{\pi}{2}$ case, by
focusing on the particular value $\delta=0$ (the qualitative
features are the same for all other values). As shown in
Fig.\,\ref{figSG}, as soon as a $\lambda\neq 0$ in switched on in
the potential, the degeneracy between neighbouring vacua of the
Sine--Gordon model is spoiled, with a consequent confinement of
the solitons. At the same time, a new static kink appears, which
interpolates between the vacua at $\phi=0$ and
$\phi=\frac{4\pi}{\beta}$ in Fig.\,\ref{figSG}. The semiclassical
method can be applied to this excitation to determine the
corresponding neutral bound states. Complementarily, FFPT is
capable of estimating the corrections to the masses of
Sine--Gordon breathers which, being non--topological excitations,
are not confined as the corresponding solitons. Therefore, the two
techniques combine together in the analysis of the full spectrum
of the model.

A qualitatively different scenario appears at the value
$\delta=\frac{\pi}{2}$. In this case, the degeneracy of two
neighbouring minima of the Sine--Gordon model is not destroyed,
hence the solitons are not confined but simply deformed into a
"large kink" and a "small kink" which interpolate between the
minima separated by a larger or smaller wall, respectively (see
Fig.\,\ref{figSG}). In this case, a straightforward application of
the semiclassical method leads to wrong results. It seems that
there are two towers of neutral states with different masses
around each vacuum, one obtained as bound states of long kink and
long antikink, and the other from short kink and short antikink.
This is in contradiction with the fact that the breathers of the
unperturbed Sine--Gordon model are non degenerate, and it is
moreover disproved by numerical analysis\cite{hungnumeric}. The
controversy has been clarified in Ref.\cite{giuseppe}, by noticing
that, at leading order in the coupling in which the semiclassical
form factor (\ref{ffinf}) is computed, the short and long kink are
invisible to each other, while the correct spectrum of neutral
states must be the result of the interaction between the two
different kinks. An exact formula for the masses of neutral states
is still unknown, yet the semiclassical method provides useful
information for the limiting cases when the masses of long and
short kink are very close, or when the large kink is much heavier
than the small (see Ref.\cite{giuseppe} for details).

The problem outlined above is typical of every potential where
kinks of different masses originate from the same vacuum.
Therefore, particular care has to be adopted in applying the
semiclassical method to those cases. For further comments and
developments, see the discussion in Ref.\cite{giuseppe}.

\section{Finite--size effects}\label{sectfinitesize}

Quantum field theory on a finite size is a subject of both
theoretical and practical interest. Besides providing a way to
control the extrapolation procedure of numerical simulations, the
study of a theory in finite volume also permits to follow the
renormalization group flow between the ultraviolet (UV) conformal
limit and the infrared (IR) massive behaviour. Clearly understood
in CFT (see Ref.\cite{finitesize}), finite--size effects can be
tackled non--perturbatively in integrable QFT as well, with the
so--called Thermodynamic Bethe Ansatz method\cite{TBA}, which is a
combination of analytical and numerical procedures.

In this Section, we will discuss the contributes to this subject
obtained through the semiclassical
method\cite{finvolff,SGscaling,SGstrip,phi4}. Once the proper
classical solutions are identified to describe a given geometry,
the spectral function in finite volume can be easily estimated by
adapting the Goldstone and Jackiw's result. Furthermore, the
finite--volume kinks can be quantized semiclassically via the DHN
technique, which permits to write in analytic form the discrete
energy levels as functions of the size of the system.

We will now explicitly show the construction in the case of the
Sine--Gordon model, defined by the potential (\ref{VSG})
\begin{equation*}
V(\phi)\,=\,\frac{m^2}{\beta^{2}}\,(1-\cos\beta\phi)\;.
\end{equation*}
We chose this integrable theory as the guiding example in this
Section, since the analysis of its finite size effects is
technically simpler than in the $\phi^4$ theory. Full details
about the $\phi^4$ model can be found in
Refs.\cite{finvolff,phi4}. Being an integrable theory, the
Sine--Gordon model has been already studied on a finite size by
appropriate extensions of the Thermodynamic Bethe
ansatz\cite{SGTBA}. In comparison with those techniques, the
semiclassical method provides more explicit anlytical results. It
would be interesting to perform a quantitative comparison between
the two approaches, in order to directly control the range of
validity of the semiclassical approximation, as it was done in the
infinite volume case.

\subsection{Classical solutions and form factors}

\begin{figure}[ht]
\begin{tabular}{p{8cm}p{8cm}}

\footnotesize

\psfrag{phicl(x)}{$\beta\phi^{+}_{cl}$} \psfrag{2 pi}{$2\pi$}
\psfrag{K(k^2)}{$\textbf{K}(k^{2})$}
\psfrag{-K(k^2)}{$-\textbf{K}(k^{2})$}
\psfrag{x}{$\hspace{0.2cm}mx/k$}

\psfig{figure=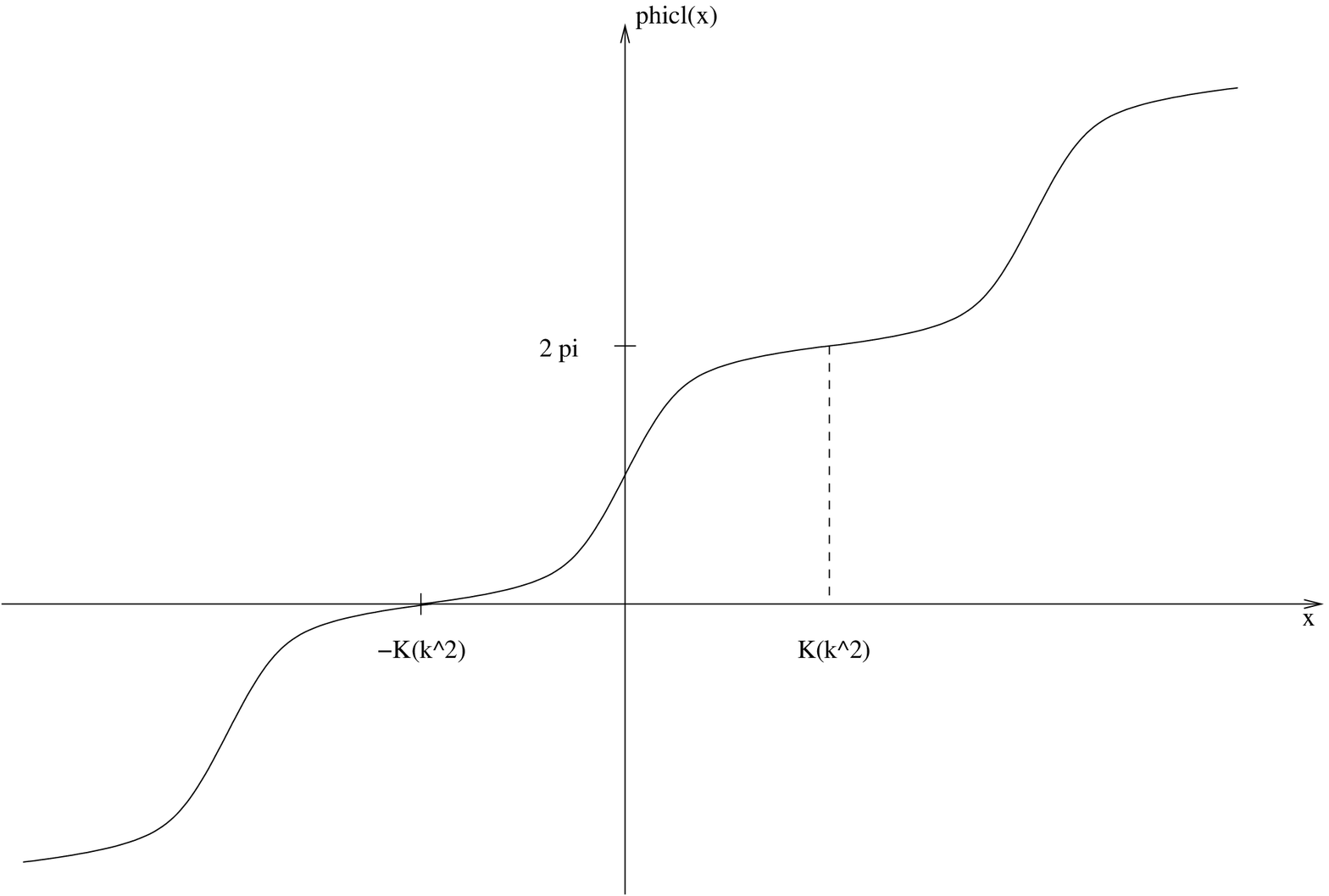,height=4cm,width=6cm}&

\footnotesize

\psfrag{phicl(x)}{$\beta\phi^{-}_{cl}$}
\psfrag{phi0}{$\beta\tilde{\phi}$} \psfrag{2
pi-phi0}{$2\pi-\beta\tilde{\phi}$}
\psfrag{K(k^2)}{$\textbf{K}(k^{2})$}
\psfrag{-K(k^2)}{$-\textbf{K}(k^{2})$}
\psfrag{x}{$\hspace{0.2cm}mx$}

\psfig{figure=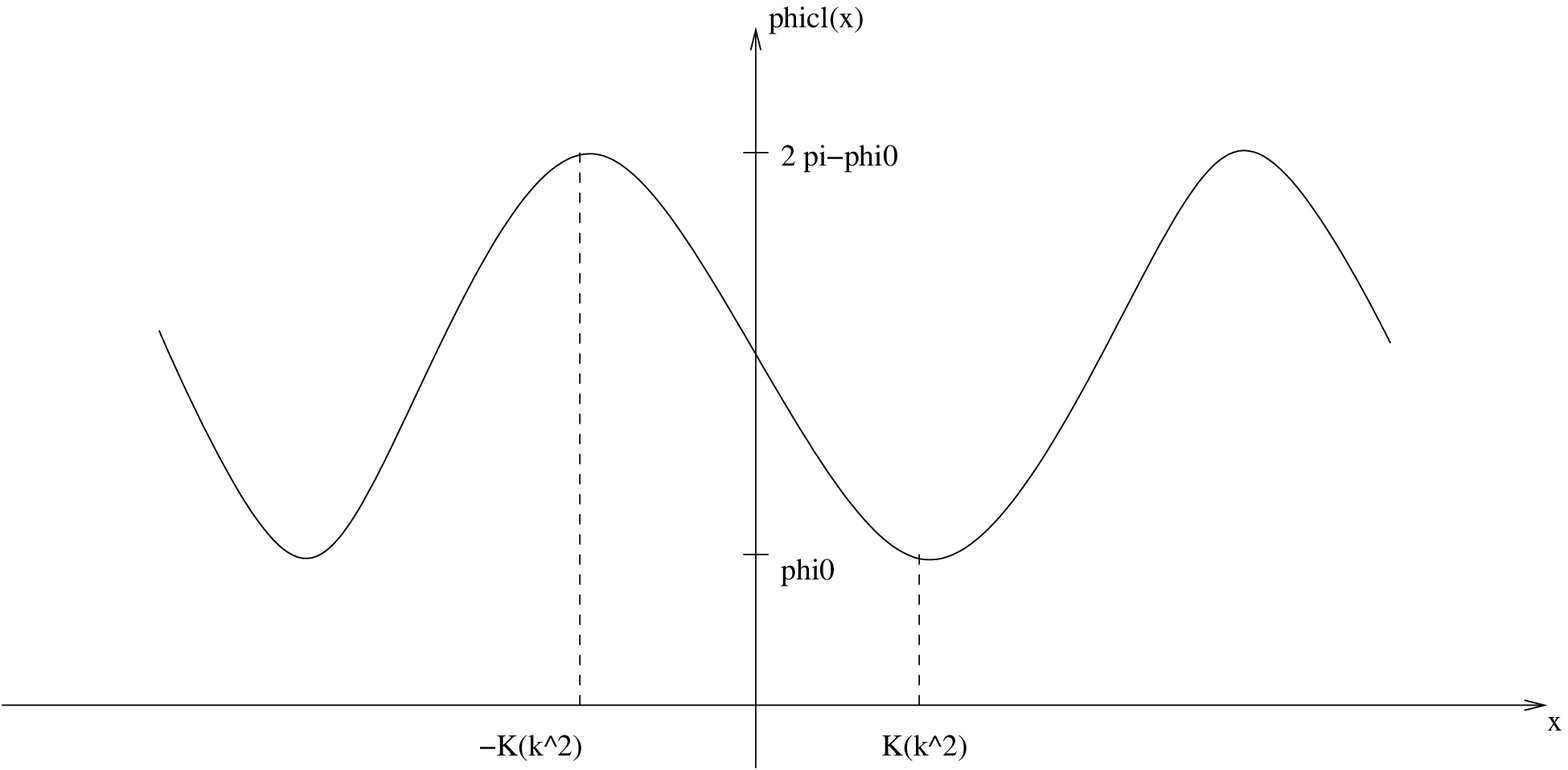,height=4cm,width=6cm}
\end{tabular}
\caption{Solutions of eq.\,(\ref{firstorder}), $A > 0$ (left hand
side), $-2 < A < 0$ (right hand side).}
 \label{figSGsol}
\end{figure}

The basic ingredient in the semiclassical study of finite size
effects is the classical kink solution on a finite volume. This
can be obtained by solving eq.\,(\ref{firstorder}) with an
appropriate constant $A$ to encode the chosen boundary conditions.
We will now focus on a cylindrical geometry of circumference $R$,
where the b.c. for a single kink can be quasi--periodic or
antiperiodic:
\begin{equation}\label{bc}
\phi(x+R)\,=\,\frac{2\pi}{\beta}\pm \phi(x)\;,
\end{equation}
and correspond to $A>0$ and $-2\frac{\beta^2}{m^2}<A<0$ in
eq.\,(\ref{firstorder}), respectively. The associated classical
solutions are expressed in terms of elliptic
functions\footnote{For definitions and properties of elliptic
integrals and Jacobi elliptic functions, see Ref.\cite{GRA}},
whose modulus $k$ is related to the size $R$:
\begin{eqnarray}\label{SGkinkfinvol}
&\phi_{cl}^{+}(x)\,=\,\frac{\pi}{\beta}+\frac{2}{\beta}\,\text{am}\left(\frac{mx}{k},k^{2}\right)\;,\quad
&k^2=\frac{2}{2+\frac{m^2}{\beta^2}A}\;,\quad R=\frac{2}{m}\,k\,\textbf{K}(k^2)\\
&\phi_{cl}^{-}(x)\,=\,\frac{2}{\beta}\,\arccos\left[\,k\,\text{sn}\left(mx,k^{2}\right)\right]\;,\quad
&k^2=\frac{\frac{m^2}{\beta^{2}}A+2}{2}\;,\quad
R=\frac{2}{m}\,\textbf{K}(k^2)
\end{eqnarray}
(see Fig.\,\ref{figSGsol}).

For simplicity, we will now only discuss the quasi--periodic case
$\phi_{cl}^{+}$, which is characterized by a classical energy
\begin{equation}\label{classen}
{\cal
E}_{cl}(R)\,=\,\frac{8m}{\beta^2}\left[\frac{\textbf{E}(k^2)}{k}+\frac{k}{2}\left(1-\frac{1}{k^2}\right)
\textbf{K}(k^2)\right]\;.
\end{equation}
A complete treatment of the antiperiodic background
$\phi_{cl}^{-}$ can be found in Ref.\cite{finvolff,phi4}, while
the study of other backgrounds related to a strip geometry with
Dirichlet boundary conditions has been performed in
Ref.\cite{SGstrip}.

We will first show that the relation between classical solutions
and semiclassical form factors presented in
Section\,\ref{secsemFF} also holds in finite volume. This comes
from the possibility of choosing $\hat{f}(a)$ as a solution of
eq.\,(\ref{firstorder}) with any constant $A$, which is related to
the size of the system. We have now to consider the matrix
elements of $\phi(0)$ between two eigenstates $|p_{n_{1}}\rangle$
and $|p_{n_{2}}\rangle$ of the finite volume hamiltonian $H_R$.
These states can be naturally labelled with the so-called
"quasi-momentum" variable $p_{n}$, which corresponds to the
eigenvalues of the translation operator on the cylinder (multiples
of $\pi/R$), and appears in the space dependent part of
eq.\,(\ref{heis}) in the case of finite volume\footnote{For large
$R$, the quasi-momentum is related to the free momentum
$p^{\infty}$ of the infinite volume asymptotic states by the
so--called Bethe ansatz equation
$p_{n}^{\infty}+\frac{\delta(p_{n}^{\infty})}{R}=\frac{2n\pi}{R}\equiv
p_{n}$, where $\delta(p^{\infty})$ is a phase shift which encodes
the information about the interaction.}. Defining $\theta_{n}$ as
the "quasi-rapidity" of the kink states by
\begin{equation*}
p_{n}=M(R)\sinh\theta_{n}\simeq M(R)\theta_{n}\;,
\end{equation*}
we can now write the form factor at a finite volume by replacing
the Fourier integral transform with a Fourier series expansion:
\begin{equation}\label{ff}
f(\theta_{n})\,=\,\langle
p_{n_{2}}|\,\phi(0)\,|p_{n_{1}}\rangle\,=\,
M(R)\int\limits_{-R/2}^{R/2} da\,e^{i\,M(R)\theta_{n}
a}\phi_{cl}(a)\;,
\end{equation}
where
\begin{equation*}
M(R)\,\theta_{n}\,\simeq\,
p_{n_{1}}-p_{n_{2}}\,=\,\frac{(2n_{1}-1)\pi}{R}\, -\,
\frac{(2n_{2}-1)\pi}{R}\,\equiv\,\frac{2n\pi}{R}\;.
\end{equation*}
This result, of very general applicability, adds to previous
studies of finite volume form factors\cite{refsff}, which on the
contrary deeply rely on the integrable structure of the considered
models. In our particular example, the Fourier transform of
(\ref{SGkinkfinvol}) gives
\begin{eqnarray*}
f(\theta_{n})&=&\frac{2\pi}{\beta}\left\{\frac{M}{2}\,R\,
\delta_{M\theta_{n},0}-i\,\frac{1-\delta_{M\theta_{n},0}}{\theta_{n}}
\left[\cos\left(M\theta_{n}\,R/2\right)-
\frac{\sin\left(M\theta_{n}\,R/2\right)}{M\theta_{n}R/2}\right]+\right.\\
&&\hspace{0.8cm}+\left.i\,\frac{1}{\,\theta_{n}\,\cosh\left(k\,\textbf{K}'\frac{M}{m}\,
\theta_{n}\right)}\right\}\;,
\end{eqnarray*}
where the kink mass $M$ can be approximated at leading order with
its classical energy (\ref{classen}).

\subsection{Energy levels}

We will now briefly sketch the semiclassical derivation of energy
levels in finite volume (for a detailed discussion see
Ref.\cite{SGscaling}). The aim is to obtain analytical expressions
for the energies $E_i(R)$ as functions of the circumference $R$.
The procedure consists in adapting the DHN quantization, outlined
in Sect.\,\ref{sectsemicl}, to the finite geometry.

\begin{figure}[ht]
\begin{center}
\psfig{figure=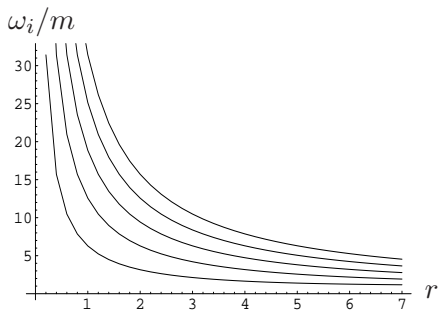,height=4cm,width=6cm} \caption{The first
few levels defined in (\ref{omegan})}\label{figomegai}
\end{center}
\end{figure}

We already know the explicit expression of the classical kink
(\ref{SGkinkfinvol}) satisfying quasi--periodic boundary
conditions. In order to construct the scaling functions, we have
to solve the corresponding Schr\"{o}dinger equation
(\ref{stability}) and to derive an analytical expression for its
frequencies $\omega_k$. Here we will not discuss the mathematical
details of this procedure, which is explained in
Ref.\cite{SGscaling}; in essence, the stability equation turns out
to be of the so--called $N=1$ Lam\'e type, for which exact
solutions are known in terms of elliptic and Weierstrass
functions. The final result for the frequencies is
\begin{equation}\label{omegan}
\omega_n^2\,=\,\frac{m^2}{k^2}\,\left[\frac{2-k^2}{3}-{\cal
P}(iy_n)\right]\;,
\end{equation}
where $y_n$ is defined by
\begin{equation}
2\textbf{K}i\,\zeta(iy_n)+2 y_n \zeta(\textbf{K})\,=\,2n\pi\;,
\end{equation}
which has the physical meaning of momentum quantization (see
Ref.\cite{GRA} for definitions of the Weierstrass functions ${\cal
P}$ and $\zeta$). By inverting the relation between $R$ and $k$ in
(\ref{SGkinkfinvol}), it is easy to plot the frequencies
(\ref{omegan}) (see Fig.\,\ref{figomegai}), which represent the
energies of the excited states (\ref{tower}) with respect to the
ground state (\ref{e0}).

\psfrag{omega1}{$\omega_{i}/m$}\psfrag{ell}{$r$}

In order to derive the ground state energy in the kink sector, it
is necessary to regularize the infinite sum (\ref{e0}), by
subtracting to it the ground state energy in the vacuum sector and
an appropriate mass counterterm. The procedure can be explicitly
carried out in the ultraviolet (UV) and infrared (IR) regimes,
when $r=mR\to 0$ and $r\to\infty$, respectively. It is
particularly interesting to compare the corresponding limiting
behaviours with asymptotic results already known for the
Sine--Gordon theory.

The small $r$ expansion is
\begin{equation}\label{grstatefinalexp}
\frac{E^{\text{kink}}_{0}(R)-E^{\text{vac}}_{0}(R)}{m} \,=\,
\,\frac{2\pi}{r}\,\frac{\pi}{\beta^{2}}\,+
\,\frac{1}{\beta^{2}}\,r
\,-\,\frac{1}{8}\,\left(\frac{r}{2\pi}\right)^{2}\,-\,
\left(\frac{r}{2\pi}\right)^{3} \left[\frac{1}{8}\, \zeta(3)
-\frac{1}{4} (2\,\log 2 -1) - \frac{\pi}{2\beta^{2}}\right] +
\ldots\;,
\end{equation}
and it has to be compared with the Conformal Field Theory
prediction\cite{finitesize}
\begin{equation}
E_0(R)\,=\,\frac{2\pi}{R}\left(\Delta_0-\frac{c}{12}\right)\,+\,BR\,+\,...\;,
\end{equation}
where $c$ is the central charge, $\Delta_0$ is the lowest scaling
dimension in the sector under consideration, and $B$ is the
so--called bulk coefficient. For the Sine--Gordon model the bulk
energy term is given by\cite{bulkterm}
$$
B\,=\,16\,\frac{m^2}{\gamma^2}\,\tan\frac{\gamma}{16}\;,\qquad\text{with}\quad
\gamma=\frac{\beta^2}{1-\frac{\beta^2}{8\pi}}\;,
$$
hence the corresponding term in (\ref{grstatefinalexp}) has the
correct small--$\beta$ behaviour. Moreover, the scaling dimension
of the kink--creating operator is known to be\cite{BerLec}
$$
\Delta_0\,=\,\frac{\pi}{\beta^2}\;,
$$
again in agreement with (\ref{grstatefinalexp})\footnote{The
central charge $c$ does not appear in (\ref{grstatefinalexp}),
since it cancels out in the subtraction of the ground state energy
in the vacuum sector}.

We should now look at the IR limit of the kink energy, and compare
it with the asymptotic approach to the infinite volume kink mass
predicted by L\"uscher's theory\cite{luscher}:
$$
M(R)-M(\infty)\,=\,2\,m\,\sin\left(\frac{\pi}{2}+\frac{\gamma}{16}\right)\,\cot\frac{\gamma}{16}\,
e^{-m\sin\left(\frac{\pi}{2}+\frac{\gamma}{16}\right)R}\,+\,O(e^{-2mR})\;.
$$
At leading order in $\beta$, this behaviour can be already
detected at the level of the classical energy (\ref{classen}),
whose IR expansion is
$$
{\cal
E}_{cl}(R)\,=\,\frac{8m}{\beta^2}\,+\,m\,\frac{32}{\beta^2}\,e^{-mR}\,+\,O(e^{-2mR})\;,
$$
where the kink mass in infinite volume is
$M_{\infty}=\frac{8m}{\gamma}$.

The successful check with known UV and IR asymptotic behaviours
confirms the ability of the semiclassical method to analytically
describe the scaling functions of SG model in the one--kink
sector.

\section{Conclusions}\label{conclusions}

We have briefly reviewed some fruitful applications of
semiclassical methods to the study of non--integrable QFT and
finite--size effects in two dimensions.

In order to keep the discussion reasonably short, we have omitted
to mention some interesting phenomena which can be captured by the
semiclassical method, like unstable resonance states and false
vacuum decay. Details can be found in the original literature.

Let us mention two of the several open problems which deserve
further attention. First, the study of semiclassical form factors
at higher order in the coupling constant would lead not only to
quantitatively better results, but also to a satisfactory
understanding of the spectrum in models where two different kinks
emerge from the same vacuum, as we discussed in the Double
Sine--Gordon case. Second, a systematic investigation of the
finite--size spectrum beyond the one--kink sector is still
missing. This requires to find appropriate time--dependent
classical solutions on a finite size and apply to them the
semiclassical quantization procedure.

\section*{Acknowledgments}

I am extremely grateful to Giuseppe Mussardo and Galen Sotkov for
their collaboration on the whole project outlined in this review.
I thank Gesualdo Delfino for many explanations and discussions,
which greatly contributed to the project besides his direct
involvement in the most recent paper.

The author's work is supported by the Giorgio and Elena Petronio
Fellowship Fund, and by the National Science Foundation under
agreement No. DMS-0111298. Any opinions, findings and conclusions
or recommendations expressed in this material are those of the
author and do not necessarily reflect the views of the National
Science Foundation.

I started to write this review when I was a postdoctoral research
assistant at the Rudolf Peierls Centre for Theoretical Physics,
University of Oxford, funded by EPSRC under the grant
GR/R83712/01, whose support I gratefully acknowledge.


\begin{thebibliography}{0}





\bibitem{CFTbooks} P. Di Francesco, P. Mathieu and D. S\'en\'echal, \textit{Conformal
field theory}, Springer, New York (1997).

\bibitem{gmrep} G. Mussardo, Phys. Rep. 218 (1992) 215, and
references therein.







\bibitem{DHN} R.F. Dashen, B. Hasslacher and A. Neveu, Phys. Rev. D 10
(1974) 4130; Phys. Rev. D 11 (1975) 3424.


\bibitem{GJ} J. Goldstone and R. Jackiw, Phys.Rev.D 11 (1975) 1486.


\bibitem{finvolff} G. Mussardo, V. Riva and G. Sotkov,
Nucl. Phys. B 670 (2003) 464.

\bibitem{dsgmrs} G. Mussardo, V. Riva and G. Sotkov,
Nucl. Phys. B 687 (2004) 189.

\bibitem{SGscaling} G. Mussardo, V. Riva and G. Sotkov,
Nucl. Phys. B 699 (2004) 545.


\bibitem{SGstrip} G. Mussardo, V. Riva and G. Sotkov,
Nucl. Phys. B 705 (2005) 548.

\bibitem{phi4} G. Mussardo, V. Riva, G. Sotkov and G. Delfino, Nucl. Phys. B 736 (2006)
259.

\bibitem{giuseppe} G. Mussardo, \textit{Neutral bound states in kink--like
theories}, hep-th/0607025.

\bibitem{raj} R. Rajaraman, \textit{Solitons and instantons},
Amsterdam, North Holland, 1982.





\bibitem{DMS} G. Delfino, G. Mussardo and P. Simonetti,
Nucl. Phys. B 473 (1996) 469.

\bibitem{SGintegrable} A.B. Zamolodchikov and Al.B. Zamolodchikov,
Ann. Phys. 120 (1979) 253; M. Karovski and P. Weisz, Nucl. Phys. B
139 (1978) 445.

\bibitem{dm} G. Delfino and G. Mussardo, Nucl. Phys. B 516 (1998) 675.

\bibitem{hungDSG} Z. Bajnok, L. Palla, G. Takacs and F. Wagner,
Nucl. Phys. B 601 (2001) 503.

\bibitem{hungnumeric} G. Takacs and F. Wagner, Nucl. Phys. B 741
(2006) 353.







\bibitem{finitesize} \textit{Finite--Size Scaling}, edited by J.L. Cardy,
Amsterdam, North--Holland, 1988.

\bibitem{TBA} Al.B. Zamolodchikov, Nucl.Phys. B 342 (1990) 695.

\bibitem{SGTBA} C. Destri and H.J. de Vega, Phys. Rev. Lett. 69 (1992) 2313;
Nucl. Phys. B 438 (1995) 413; D. Fioravanti, A. Mariottini, E.
Quattrini and F. Ravanini, Phys. Lett. B 390 (1997) 243; G.
Feverati, F. Ravanini and G. Takacs, Phys. Lett. B 430 (1998) 264;
Nucl. Phys. B 540 (1999) 543.

\bibitem{GRA} I.S. Gradshteyn and I.M. Ryzhik,
\textit{Table of integrals, series, and products}, Academic Press,
New York (1980); E.T.Whittaker and G.N.Watson, \textit{A course of
modern analysis}, Cambridge, Cambridge University Press, 1927.


\bibitem{refsff} F. Smirnov, Amer. Math. Soc. Transl. (2), Vol. 201 (2000),
283; P. Fonseca and A. Zamolodchikov, J. Stat. Phys. 110 (2003)
527; A.I. Bugrij, \textit{Form factor representation of the
correlation function of the two dimensional Ising model on a
cylinder}, hep-th/0107117; V.E. Korepin and N.A. Slavnov, Int. J.
Mod. Phys. B 13 (1999) 2933.




\bibitem{bulkterm} Al.B. Zamolodchikov, Int. J. Mod. Phys. A 10 (1995) 1125.

\bibitem{BerLec} D. Bernard and A. LeClair, Comm. Math. Phys. 142 (1991)
99.

\bibitem{luscher} M. L\"{u}scher, Comm. Math. Phys 104 (1986) 177;
T.R. Klassen and E. Melzer, Nucl. Phys. B 362 (1991) 329.








\end{thebibliography}
\end{document}